\documentclass[]{aastex631}

\usepackage{CJKutf8}
\usepackage{multirow}
\usepackage{amssymb}
\makeatletter

\newcommand{\Rmnum}[1]{\expandafter\@slowromancap\romannumeral #1@}
\makeatother

\begin{document}
\begin{CJK*}{UTF8}{bkai}

\title{The Detectability of Coronal Mass Ejections in the Low Corona Using Multi-slit Extreme-ultraviolet Spectroscopy}

\correspondingauthor{Xianyong Bai}
\email{xybai@bao.ac.cn}

\author[0000-0002-1943-8526]{Lami Chan(陳霖誼)}
\affiliation{School of Earth and Space Sciences, Peking University, Beijing, 100871, People’s Republic of China}

\author{Xianyong Bai}
\affiliation{National Astronomical Observatories, Chinese Academy of Sciences, Beijing 100101, China}
\affiliation{School of Astronomy and Space Sciences, University of Chinese
Academy of Sciences, Beijing, 100049, China}
\affiliation{Institute for Frontiers in Astronomy and Astrophysics, Beijing Normal University, Beijing, 102206, China}

\author[0000-0002-1369-1758]{Hui Tian}
\affiliation{School of Earth and Space Sciences, Peking University, Beijing, 100871, People’s Republic of China}

\author{Yufei Feng}
\affiliation{National Astronomical Observatories, Chinese Academy of Sciences, Beijing 100101, China}

\author[0000-0002-7421-4701]{Yu Xu}
\affiliation{School of Earth and Space Sciences, Peking University, Beijing, 100871, People’s Republic of China}

\author[0000-0003-3843-3242]{Tibor T\"{o}r\"{o}k}
\affiliation{Predictive Science Inc., 9990 Mesa Rim Road, Suite 170, San Diego, CA 92121, USA}

\author[0000-0002-6641-8034]{Yuhang Gao}
\affiliation{School of Earth and Space Sciences, Peking University, Beijing, 100871, People’s Republic of China}

\author[0000-0002-9667-6392]{Tanmoy Samanta}
\affiliation{Indian Institute of Astrophysics, Kormangala, Bangalore, INDIA}

\author[0000-0001-5657-7587]{Zheng Sun}
\affiliation{School of Earth and Space Sciences, Peking University, Beijing, 100871, People’s Republic of China}




\begin{abstract}

The spectra of coronal mass ejections (CMEs) in the low corona play a crucial role in understanding their origins and physical mechanism, and enhancing space weather forecasting. However, capturing these spectra faces significant challenges. This paper introduces a scheme of a multi-slit spectrometer design with five slits, acquiring the global spectra of the solar corona simultaneously with a focus on the spectra of CMEs in the low corona. The chosen wavelength range of the spectrometer (170-180 $\mathrm{\AA}$) includes four extreme ultraviolet emission lines (Fe~{\sc{x}} 174.53 $\rm \AA$, Fe~{\sc{ix}} 171.07 $\rm \AA$, Fe~{\sc{x}} 175.26 $\rm \AA$, Fe~{\sc{x}} 177.24 $\rm \AA$), which provides information of plasma velocity, density and temperature. Utilizing a numerical simulation of the global corona for both the on-disk and the off-limb scenarios, we focus on resolving the ambiguity associated with various Doppler velocity components of CMEs, particularly for a fast CME in the low corona. A new application of {our} decomposition technique is adopted, enabling the successful identification of multiple discrete CME velocity components. Our findings demonstrate a strong correlation between the synthetic model spectra and the inverted results, indicating the robustness of {our} decomposition method and its significant potential for global monitoring of the solar corona, including CMEs.

\end{abstract}


\section{Introduction} \label{sec:intro}
Coronal mass ejections (CMEs) are significant solar eruption phenomena that release substantial energy and plasma into interplanetary space. These events can drive various space weather effects on Earth, which can cause damage for the electromagnetic infrastructure. Consequently, accurate space weather forecasting is crucial, necessitating an improved understanding of CMEs. However, the physical mechanisms governing their origin and evolution (especially in the low corona) remain unclear, and the accuracy of space weather forecasts is not sufficient. Spectroscopic observations of CMEs can provide important information (e.g., the line-of-sight (LOS) velocity), to enhance our understanding of CMEs and improve space weather forecasts \citep[][]{tian2012,Golub2020}. 

Current state-of-the-art technologies and instruments are not yet capable of capturing the plasma information of CMEs efficiently. Using a typical spectrometer to capture CME spectra in the corona within $\sim$1.5 $R_{\odot}$ (hereafter CME spectra) is challenging because CMEs can appear at arbitrary locations on the Sun. The traditional single-slit spectrometers, such as the extreme ultraviolet (EUV) Imaging Spectrometer \citep[EIS,][]{Culhane2007} on board Hinode, is compromised by their limited field of view (FOV) and low cadence, resulting in meager observations of CME spectra \citep[e.g.,][]{tian2012}. The transient and large-scale features of CMEs necessitate the implementation of global EUV spectroscopic observations with high temporal resolution to effectively monitor the corona, facilitating the capture of CME spectra. 
The Extreme Ultraviolet Variability Experiment \citep[EVE,][]{Woods2012} on board the Solar Dynamics Observatory (SDO) provides full-disk spectroscopic observations and has been used to acquire Doppler velocity of CMEs but lacks spatial resolution \citep[e.g.,][]{Xu2022,Lu2023}. The Coronal Multichannel Polarimeter \citep[CoMP,][]{Tomczyk2008} can obtain Doppler shift and line width of CMEs using near-infrared (NIR) spectroscopy, but it is limited in observing the off-limb corona \citep[e.g.,][]{Tian2013}. 
The Ultraviolet Coronagraph Spectrometer \citep[UVCS,][]{Kohl1995} on board the Solar and Heliospheric Observatory \citep[SOHO,][]{Delaboudiniere1995}, utilizes far-ultraviolet spectroscopy to analyze CMEs, with a focus on the higher corona, extending up to $12R_{\bigodot}$. 
{The Hinode/EIS slot observations, and the Multi-Order Solar EUV Spectrograph \citep[MOSES,][]{Fox2010} sounding rocket experiment, provide extended (instantaneous) field-of-view spectroscopic observations.}
Similarly, the Visible Emission Line Coronagraph \citep[VELC,][]{Patel2021} on board Aditya-L1 is capable of investigating CME spectra using multi-slit observations at visible and NIR wavelengths (Fe~{\sc{xiv}} 5303 $\rm \AA$, Fe~{\sc{xi}} 7892 $\rm \AA$ and Fe~{\sc{xiii}} 10747 $\rm \AA$), but is limited by the absence of on-disk observations \citep[][]{Ramesh2024}.

A novel approach in EUV spectroscopy involves the use of a multi-slit design, facilitating high-cadence observations. However, this introduces the multi-slit ambiguity, which refers to the {spectral} overlapping from different slits in \citet[][]{Cheung2019}. To address this ambiguity, a newly developed spectral decomposition technique has been applied to separate the overlapping spectra from different slits. This technique incorporates key strategies from the differential emission measure (DEM) inversion technique, which have been extensively developed over time \citep[see a review in][]{DelZanna2018}. {Our decomposition technique also follows the key concept of velocity DEMs (VDEMs) in \cite{Cheung2019}.} \citet{Cheung2015} employed an inversion method with sparse solutions for emission measure (EM), using observations from the Atmospheric Imaging Assembly \citep[AIA,][]{Lemen2012} on board SDO. However, this method only considered temperature in the parameter space and used the intensities of six EUV channels, without incorporating spectral information. 
\citet{Cheung2019} described a general framework for a decomposition technique applicable to both single-slit (e.g., \textit{Hinode}/EIS), and multi-slit instruments, particularly for the Multi-slit Solar Explorer \citep[MUSE,][]{DePontieu2020}. As a proposed mission to be launched in 2027, MUSE will employ an advanced 37-slit design and the decomposition technique to resolve overlapping spectra, thereby enhancing our understanding of coronal heating through spectroscopic observations of Doppler shifts and line widths. This framework is also applicable to slitless instruments such as the COronal Spectroscopic Imager in the EUV \citep[COSIE,][]{Winebarger2019}, a proposed mission, and the Marshall Grazing Incidence X-Ray Spectrometer \citep[MaGIXS,][]{Savage2023}, for which a corresponding inversion method (herein referred as the decomposition technique) for slitless spectrograph data have been developed \citep[e.g.,][]{Athiray2024}. COSIE aims to provide high-cadence EUV spectroscopic observations on a global scale, and its applications to CME studies is planned for a future goal \citep[][]{Winebarger2019}. While COSIE has the potential to provide high-cadence observations of CME spectra, a challenge lies in the complexity of the spectral decomposition during CMEs. \citet{Chan2024} provides a scheme with five-slit design for global coronal observations, obtaining a series of plasma diagnostics. However, this scheme may face difficulties in capturing the spectra of fast CMEs due to its relatively crowded wavelength range (184-197 Å). 

In this paper, we presents a scheme of a five-slit EUV spectrograph with a new application of {our} decomposition technique to resolve velocity ambiguity, optimizing for the detectability of CME spectra in the low corona. This allows us to not only monitor the quiet corona (global velocity and density diagnostics), but also extract substantially robust CME spectra (velocity and partial density diagnostics), particularly for a fast CME with discrete multiple Doppler velocity components.
Section~\ref{sec:description} provides a detailed description of the proposed scheme. 
Section~\ref{sec:decomposition} introduces a new application of {our} decomposition technique with specific strategies during a fast CME. Section~\ref{sec:results} presents global plasma diagnostics for both the on-disk and the off-limb scenarios of a fast CME and compares the inverted results with the synthetic model spectra from a numerical model. In Section~\ref{sec:conclusion}, we discuss our findings and draw conclusions.

\section{Scheme Description} \label{sec:description}
The multi-slit design of an EUV spectrograph for full-disk observations provides an efficient approach to probe the global solar corona, particularly for CMEs. We propose a preliminary scheme of an EUV spectrometer with five slits within the wavelength range from 170 to 180 Å, focusing on the detectability of CME spectra in the on-disk and the off-limb scenarios. This is promising for providing invaluable insights into the origin of CMEs (particularly in the low corona), and for determining velocity vectors by combining LOS velocities (by EUV spectroscopy) with the plane-of-sky (POS) velocities (by a context imager). 
In practice, this design can be still employed to investigate the quiet Sun (QS) and active regions (ARs) when the Sun is not so active, e.g., the origin of solar wind by monitoring coronal holes (CHs) on the Sun. During solar eruptions like CMEs, it is capable of capturing CME spectra efficiently and obtaining their plasma information while the potential schemes for routine detection of CME spectra, have yet to be achieved \citep[][]{Winebarger2019,Chan2024}

\begin{figure}[ht!]
\plotone{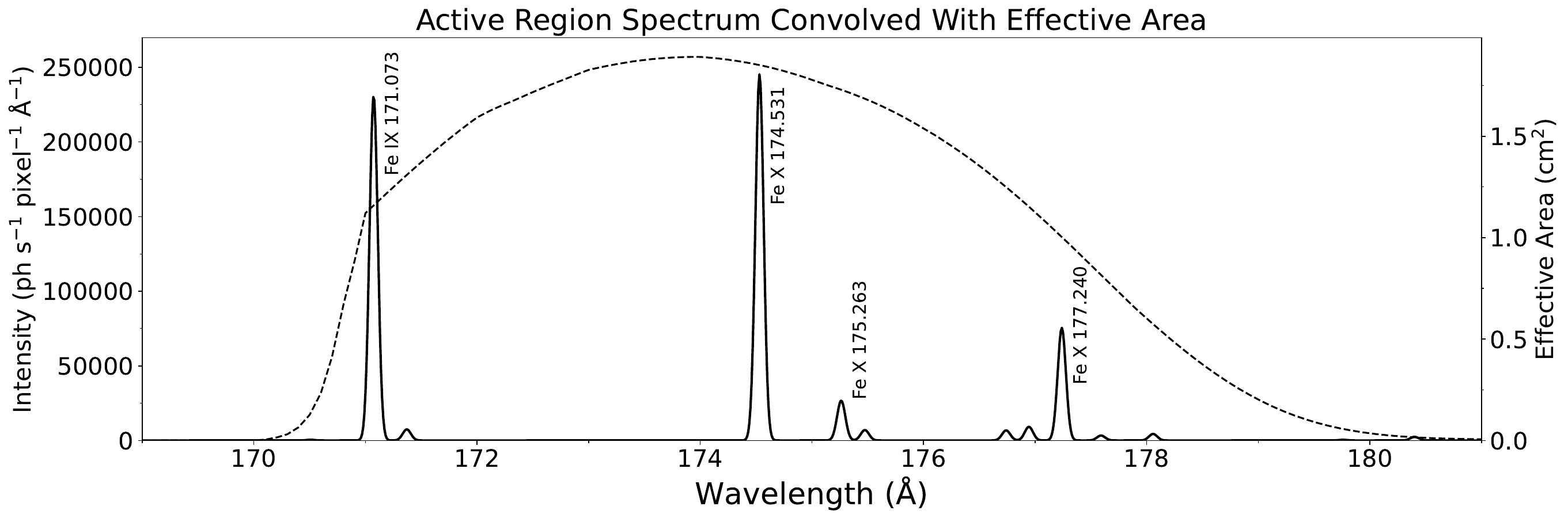}
\caption{Solar example spectrum in the target wavelength range (i.e. 170 - 180 Å) with {lines of interest} marked in black (i.e., primary lines), calculated with CHIANTI using AR DEM under an assumption of a density of $10^9\,\text{cm}^{-3}$. An effective area, {(dashed line)} with peak value of $1.89\,\text{cm}^{2}$ has been convolved with the example spectrum {(solid line)}, dropping dramatically around 170 Å.}
\label{fig:effective area}
\end{figure}

The overall instrumental parameters are largely consistent with \citet{Chan2024} and \citet{Feng2024}, given the shared mission framework and potential integration within the same instrument, albeit focusing on different wavelength ranges. Note that this is a preliminary proposed mission and we are continually refining the optimal parameters to best align with our scientific goals, resulting in some slight variations of instrumental parameters. The wavelength range in this work is 170-180 Å (vs. 184-197 Å in \citet{Chan2024} and \citet{Feng2024}). This adjustment is motivated by the less crowded spectral lines compared to those in the 184-197 Å range. Figure~\ref{fig:effective area} illustrates four isolated primary lines with an example spectrum convolved with effective area (dashed line), facilitating the identification of different velocity components with a clean wavelength range, particularly for those associated with fast CMEs (which usually manifest as discrete components). This clean wavelength range reduces potential overlap between discrete CME components and other spectral lines (primary or other CME components both are possible). Furthermore, the inter-slit spacing has been increased to 17.7 Å (vs. 1.02 Å), minimizing overlap between different slits. Consequently, the major concern now shifts to managing the overlap between discrete CME components and other contaminant lines, as well as accurately locating discrete CME components, particularly in the context of fast CMEs.
Considering the enhancement of the inter-slit spacing (17.7 Å with a corresponding slit separation in the POS of $600''$) and the spectral sampling  of $\sim$$\rm 0.04\,Å\,pixel^{-1}$,  the detector size has been expanded to $2048$ {pixels} (vs. $1024$ {pixels}). The FOV has also increased slightly to $3000''\times3000''$ (vs. $2400''\times2400''$) to investigate the upper limits of the signal detectability{. This extension allows the FOV to reach beyond 1.5 $R_{\odot}$, where signals are weak}, particularly {during non-eruption periods with lower intensities} in the off-limb scenarios. An effective area (Figure~\ref{fig:effective area}), with a slightly larger peak value of $\rm 1.89\,cm^{2}$ (vs. $\rm 1.68\,cm^{2}$), drops sharply at wavelengths shorter than 170 Å. The exposure time has been marginally reduced to $\rm 1.5\,s$ (vs. $\rm 2\,s$) due to the larger absolute intensity of 170-180 Å wavelength range compared to that of 184-197 Å wavelength range. As a result, despite a larger FOV, the cadence of the full-disk scanning can be maintained at $\sim$$\rm 300\,s$ {with a slit width of $\sim$4$''$)}. {This cadence is a result of consideration of readout time and slit moving, whereas it would be $\sim$$\rm 225\,s$ without considering these factors}. In {practical} scenarios with different observation goals, the cadence varies significantly. For instance, capturing CMEs in ARs requires an exposure time of less than $\rm 1\,s$ (e.g., $\rm 0.8\,s$), resulting in a cadence of $\sim$2 minutes without considering readout time and slit moving. This cadence is deemed sufficient for diverse phenomena, particularly for eruptive phenomena, such as flares and CMEs (reaching up to $\sim$$\rm 1.5\,R_{\odot}$), evolving {on timescales} of 10 minutes \citep[][]{Cheng2020,Tamburri2024}.

\section{The Application of Decomposition}\label{sec:decomposition}
\subsection{Our Decomposition Method and a Forward Model }
Recent advancements in three-dimensional (3D) global magnetohydrodynamic (MHD) modeling of the solar corona provide spatially resolved, full-disk plasma information (e.g., velocity, temperature, and density). 
We utilized one frame from a 3D MHD simulation of the 2000 July 14 Bastile Day eruption by Predictive Science Inc. (PSI) \citep[][]{Torok2018}, considering two viewing angles corresponding to the on-disk and the off-limb scenarios. {A magnetically stable flux rope was inserted into AR 9077, triggering the eruption through boundary-driven flows. This flux rope is shown in the subsequent discussion in the off-limb scenario}. This corresponds to the initial eruption stage of a fast CME in the low corona, reaching up to a velocity of $\sim$$\rm 3000\,km\,s^{-1}$ at $\sim$$1.4\,\rm R_{\odot}$.
Synthetic EUV spectra derived from this model (as the ground truth) and detailed description of the employed decomposition method for this scheme have been presented in \citet{Chan2024}.  Therefore, only a brief description of the decomposition theory is provided here, and we focus on a different application of {our} decomposition method in this paper. We followed the general framework of inversion described in \citet{Cheung2015,Cheung2019} and the applications \citep[][]{Winebarger2019,DePontieu2020,Savage2023,Chan2024,Athiray2024}. A clear schematic description for the decomposition process is provided in Figure 11 of \citet[][]{Savage2023}. The decomposition problem can be simplified by solving the following linear system
\begin{equation}
    \textbf{y} = \textbf{R}\textbf{x},
\label{eq:Rx}
\end{equation}
where \textbf{y}, {the synthetic observation spectrum,} is a one-dimensional array with M tuples (representing the dispersion direction of the detector). {We incorporated photon (Poisson) noise into the ground truth spectrum synthesized from the model}. In principle, \textbf{y} should contain the total spectrum of 5 slits synthesized from the model, i.e., M = 2048. However, in most practical scenarios with our design, a CME typically manifests in only one or two slits, with the rest of slits being unaffected. This means that the size of \textbf{y} can be significantly reduced to $\sim$500 or $\sim$1000  {as the input} for a CME appears in one or two slits, respectively. This dramatically reduces the computational resources and the time required for the inversion process. Note that the reduced size applies only to the input for inversion process while the detector maintains its full size of 2048 pixels.
\textbf{x}, {as the output of the equation}, is a one-dimensional array with Q tuples representing EM as a function of density, Doppler velocity, temperature and slit number. However, when a CME appears in only one or two slits, the size of Q reduces accordingly. In {our} case with one slit {(one of the input)}, \textbf{R} is the response matrix with $M\,\times\,Q$ dimensions, where $Q\,=\,Q_{\rm density}\times Q_{\rm velocity}\times Q_{\rm temperature} \approx 14000$ was constructed by finite bins of each dimension of the parameter space. Note that $Q_{\rm velocity}$ depends on the stage of CME evolution and viewing angle. For the aforementioned frame, in the on-disk (the off-limb) scenario, $Q_{\rm velocity}\,\sim\,70 (80)$ ranges from $\rm -3000\,km\,s^{-1}$ to $\rm 500\,km\,s^{-1}$ ($\rm -1300\,km\,s^{-1}$ to $\rm 2700\,km\,s^{-1}$) with a step size of $\Delta v = 50\,\text{km}\,\text{s}^{-1}$. This step size of velocity is derived by considering the resolving power of $\sim$2000 and computational efficiency, as smaller step size demands more resources. Based on the contribution function calculated from the CHIANTI atomic database v10.0.2 \citep[][]{Dere1997,DelZanna2021}, we generated a response matrix under conditions of density ($7\leq\, \mathrm{log} \,N /\text{cm}^{-3}  \leq\,12$ with $\Delta\,\mathrm{log}\,N /\text{cm}^{-3} = 0.5$), Doppler velocity (as aforementioned), temperature ($5.0\leq\,\mathrm{log}\,T/\text{K}\leq\,7.0$ with $\Delta\,\mathrm{log}\,T /\text{K} = 0.1$), and slits with a large displacement (inter-slit spacing) of 17.7 Å on the spectrogram. Finally, the output is conceptually similar to VDEMs but includes additional information about varying density. {This represents a DEM as a function of velocity, temperature, and density.} Additionally, the matrix equation is solved using a machine learning algorithm, specifically the Lasso Least Angle Regression (LassoLars) implemented in Python. Additional relevant information can be found in \citet{Chan2024}.

\subsection{Ambiguous Parameter: Velocity}
\begin{figure}[ht!]
\plotone{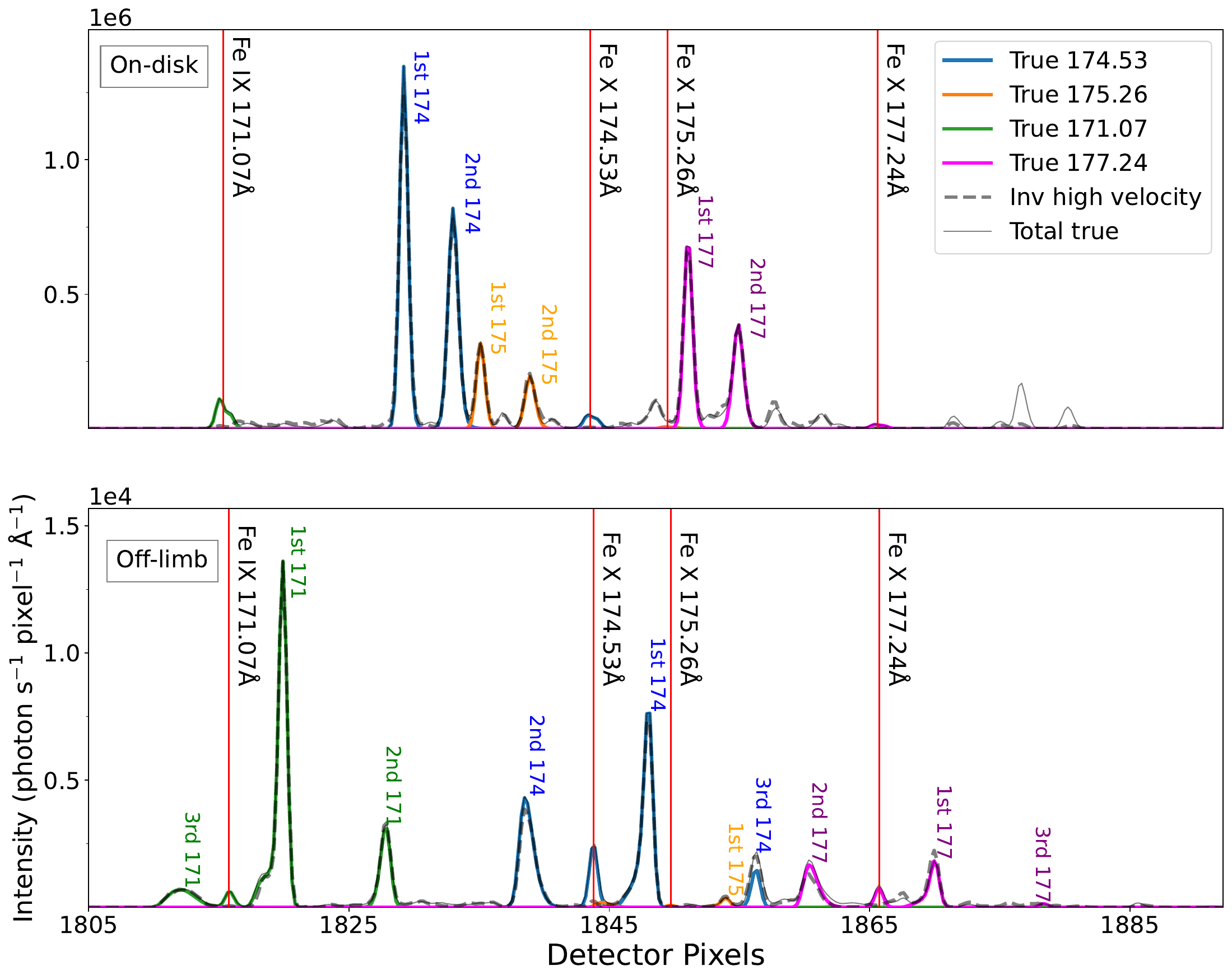}
\caption{Example spectra from two arbitrary pixels (which are marked by green circle in Figure~\ref{fig:disk_v}, Figure~\ref{fig:disk_i_n}, Figure~\ref{fig:limb_v_i}, and Figure~\ref{fig:171v}) of slit 4 in the on-disk (top panel) and the off-limb (bottom panel) scenario. Solid lines in four different color represent four primary lines, recognised by {our} decomposition technique. Their CME components are marked with corresponding numbers in the same color (e.g., ``1st 174'' means the first CME component of Fe~{\sc{x}} 174.53 $\rm \AA$). Primary components (around the rest wavelength) are marked with red vertical lines. Inverted high-velocity spectra (dashed line) generated by {our} decomposition technique are the possible CME components. The black solid line (``total true'') represents the total spectra synthesized from the ground truth (considering Poisson noise).}
\label{fig:eg_sp}
\end{figure}

In contrast to \citet{Chan2024}, where the only ambiguous parameter is the slit number, we adjusted the scheme (e.g., by changing the wavelength range and the inter-slit spacing) to reduce the number of the ambiguous parameters. If applying a CME scenario from the scheme in \citet{Chan2024}, the ambiguous parameters extend to both the slit number and velocity, significantly impacting the accuracy of the inversion process due to the increased uncertainty associated with more parameters for decomposing. For example, overlapping components of a spectral line may originate from different slits in a QS region while these components in a CME scenario may originate from different slits and velocity (CME) components (related to velocity in the parameter space). Consequently, we revised the instrumental parameters to ensure that the velocity is the only ambiguous parameter. {This improvement is achieved through two primary methods. First, we increase the inter-slit spacing from 1.02 Å in \citet{Chan2024} to 17.7 Å, significantly reducing the interaction between different slits. Second, we modify the wavelength range (from $\sim$184 Å to $\sim$197 Å) to the range of $\sim$170  Å to $\sim$180 Å, which contains fewer contaminant lines}. The subsequent challenge is to accurately decompose the different velocity components. Given the need to process large volumes of data in future analyses, it is essential to develop an automated code to efficiently and accurately identify these velocity components, which is obtained by a new application (ambiguous parameter is only velocity) of the core concept in \citet{Cheung2015}. Figure~\ref{fig:eg_sp} presents the inversion results for two sample spectra of CMEs from two arbitrary pixels in the on-disk (top panel) and the off-limb scenarios (bottom panel), respectively. These two pixels (pointed by green arrow and marked by green circle), are from slit 4, leading to considerable wavelength displacement with a inter-slit spacing of 17.7 Å on the detector. {Since we utilized only one snapshot of PSI model as the synthetic observation spectra, we visually identified the occurrence of a CME in each slit and verified it by examining the specific spectra. In actual observations, this process can be automated through an algorithm and a context imager, which will be possible in the future.} Through automated procedures, we decomposed several CME components, indicated by numbers and wavelengths (e.g., ``1st 174'' means the first CME components of Fe~{\sc{x}} 174.53 $\rm \AA$), utilizing two main strategies. For example, in the on-disk scenario, we adopted the inversion on synthetic observation spectrum (``total true'' indicated by the black solid line, {which takes into account Poisson noise and superpositions from other contaminant lines to better approach the real observations}) and can derive the spectra within each velocity bin with a velocity resolution of $\rm 50\,km\,s^{-1}$. To locate CME components with high velocities, we select only the spectrum with high-velocity components (``inv high velocity'' indicated by the dashed line) while the high-velocity spectra exhibits very low intensities for primary components as expected. {This is accomplished by using $R_1\, x_1$, which contains the elements specific to a certain slit, along with temperature and density, but only for {blueshifts} exceeding a velocity threshold} of $\sim$$\rm -400\,km\,s^{-1}$. {Velocity} in the parameter space, with blueshifts larger than this threshold classified as ``inv high velocity''. This threshold is determined by considering the smallest blueshift of a discrete CME component relative to its primary component. The line broadening is considered by incorporating thermal broadening (at coronal temperature), non-thermal broadening with a constant value $\sim$$\rm 15\,km\,s^{-1}$ under coronal conditions \citep[][]{chae1998,Sheoran2023}, and instrumental broadening (using a spectral resolving power of $\sim$$2000$). This results in a width of $\sim$$\rm 190\,km\,s^{-1}$ corresponding to the wavelength difference of 3 standard deviations (3 $\sigma$). 
For a discrete CME component to be fully separated from its primary component, the distance between their Guassian centroids should be larger than $\sim$$\rm 380\,km\,s^{-1}$.
Considering the potential enhancement of line broadening in actual observations (particularly during CMEs), a slightly larger threshold of $\sim$$\rm -400\,km\,s^{-1}$ is reasonable, e.g., see the joint probability distribution functions (JPDF) in the second row in Figure~\ref{fig:disk_v}. {It is important to note that there is limited understanding regarding the concrete values of non-thermal velocity in similar cases due to the rarity of EUV spectroscopic observations during eruptions, despite evidence of its enhancement preceding flares \citep[][]{Harra2013}. Conducting statistical analyses may be necessary to establish a reference value. From a theoretical perspective, determining a typical value during an eruption is challenging due to the complex nature of non-thermal broadening, which encompasses various factors such as MHD waves, unresolved turbulent motion, and magnetic reconnections. Furthermore, considering a resolving power of $\sim$2000 in our work, thermal broadening under coronal temperature, and non-thermal velocity speed of $\sim$$\rm 15\,km\,s^{-1}$, we estimate a typical value of line width of $\sim$$\rm 100\,km\,s^{-1}$ (as shown in \citet{Chan2024}). Given these complexities, it may be acceptable to allow for some enhancement of non-thermal velocity during eruptions in our work.}

Generating inverted spectra for high-velocity components (marked as ``inv high velocity'' in Figure~\ref{fig:eg_sp}) is the first step in rapidly and automatically identifying CME components. We employed different spectral lines from the same or closely related ions as a double-check, since the velocities derived from the same ions should be consistent. 
For example, the first CME components of Fe~{\sc{x}} 174.53 $\rm \AA$  (one of the lines of Fe~{\sc{x}}) in the on-disk scenario, are preliminarily located through {our} decomposition technique first.
We then verified the presence of similar profiles in Fe~{\sc{x}} 177.24 $\rm \AA$ and Fe~{\sc{x}} 175.26 $\rm \AA$, exhibiting the same blueshift ($\sim$$\rm 2800\,km\,s^{-1}$ in the case in Figure~\ref{fig:eg_sp}). Note that Fe~{\sc{x}} 177.24 $\rm \AA$ is primarily used, as the intensity of Fe~{\sc{x}} 175.26 $\rm \AA$ is often low and blended with other spectral lines (either primary or high-velocity components). We also included Fe~{\sc{ix}} 171.07 $\rm \AA$ in this double-check due to its similarity. However, the sharp decline in the effective area near 170 Å (Figure~\ref{fig:effective area}), with a magnitude of 0.01 $\rm cm^2$ at 170.2 Å corresponding to a blueshift of $\sim$$\rm 1500\,km\,s^{-1}$ for Fe~{\sc{ix}} 171.07 $\rm \AA$, leads to less {feasibility} for double-checking fast CME events. Additionally, uncertainties related to the instrument, {such as the manufacturing uncertainties of aluminum filter for this wavelength range}, further affect the reliability of the high-velocity components of Fe~{\sc{ix}} 171.07 Å. Consequently, in the on-disk scenario of Figure~\ref{fig:eg_sp}, there are two detected CME components for Fe~{\sc{x}} 174.53 $\rm \AA$, Fe~{\sc{x}} 175.26 $\rm \AA$, Fe~{\sc{x}} 177.24 $\rm \AA$ while Fe~{\sc{ix}} 171.07 $\rm \AA$ shows only a primary component.

Since multiple CME components are detected, and components with low intensities are less important than those with high intensities, we sorted CME components by intensity. For example, ``1st 174'' refers to the component with the largest intensity in Fe~{\sc{x}} 174.53 Å. Another possible sorting method of CME components is by velocity, however, the high-velocity components can sometimes have lower intensities than their low-velocity counterparts, which complicates the physical interpretation of plasma velocity in the LOS. For example, ``3rd 174'' in the bottom panel of Figure~\ref{fig:eg_sp}, with the largest Doppler velocity but the smallest intensity, would be ranked ``1st 174'' if sorted by velocity. \citet{tian2012} and \citet{Xu2022} show the intensities of unseparated CME components can exceed $\sim$$0.1$ of intensities of primary components. {\citet{Tian2021} shows the intensity ratio of the two components is typically between 5\% and 15\% in coronal dimming regions}. Considering that discrete CME components are likely to have lower intensities and those components with excessively low intensity yield lower confidence levels, we introduce a threshold of $0.05\,I_{\text{max}}^{l}$ to ensure the reliability of detected CME components, where $I_{\text{max}}^{l}$ is the maximum intensity among primary and CME components for each line. For example, the intensity of the first CME component of Fe~{\sc{x}} 174.53 $\rm \AA$ is considered as $I_{\text{max}}$ in Figure~\ref{fig:eg_sp}, although for low-velocity scenarios, $I_{\text{max}}^{l}$ often corresponds to the intensity of the primary component. It is worth noting that this threshold of $0.05\,I_{\text{max}}^{l}$, serves as a flexible lower limit, particularly for future real observations. The choice of this threshold value should be discussed in the context of specific real observational conditions, such as the S/N. {In addition to the components with low intensities, it is important to clarify that these components, while less significant compared to the high-velocity components, still play a crucial role. The sorting criterion helps in understanding the contribution of different velocity components to a LOS column, allowing for a quicker assessment of which velocity component is dominant. To gain a comprehensive view of the varying velocity distributions, individual maps for different CME components would be beneficial. Also, the less important components with too low intensities have already been excluded based on two criteria, which are the S/N threshold ($\rm S/N > 10$) and the proportion threshold ($0.05\,I_{\text{max}}^{l}$).}

\section{Plasma Diagnostics} \label{sec:results}
We have applied a multi-slit design and a new application of {our} decomposition technique to efficiently derive the global coronal spectra of a fast CME. We focused on Doppler shift analyses to address the ambiguities of its various velocity components. Comparison between the ground truth and inverted results for global maps will be discussed subsequently.

\subsection{On-disk CME Scenario}\label{sec:ondisk}
Figure~\ref{fig:disk_v} presents the velocity distribution in the on-disk CME scenario for the CME components (first row) and the primary components (second and third row) with markedly different velocity ranges. The maximum blueshift for the CME components (first row) reaches $\rm -3000\,km\,s^{-1}$ due to the rapid acceleration of this fast CME. The two subpanels (e.g., the ground truth in the first row) represent the first (left subpanel) and the second CME components (right subpanel) identified in the on-disk scenario. This indicates the majority of the CME components are characterized by the first components while much fewer instances reveal two distinct velocity components corresponding to two discrete CME components. The JPDF, shown in the right panel of the first row in Figure~\ref{fig:disk_v}, demonstrates a good agreement between the ground truth and the inversion results with a white dashed line showing $\rm \pm 50\,km\,s^{-1}$ uncertainties. We employed two intensity thresholds of signal-to-noise ratio ($\rm S/N = 10$) and $I_{\text{CME}} = I_{\text{max}}^{l}/20$, resulting in a small gap around $\log{I} = 2.0$. For example, a potential CME component with $\rm S/N > 10$ but $I_{\text{CME}} < I_{\text{max}}^{l}/20$ may not be classified as valid due to its low intensity, which renders it indistinguishable from other weak spectral lines. In addition, the majority of pixels are concentrated around an intensity of $\log{I} = 3$. The second row exhibits the velocity map of unseparated CME components (around primary components) by a broad velocity range compared with that of the QS region, making the velocity variations within the QS region difficult to discern. A strong redshift is observed in a belt shape surrounding the CME region (i.e., in the region representing pixels corresponding to the first CME components in the velocity map). Certain pixels show multiple CME components, including a discrete CME component, a unseparated CME component, a primary component, and a possibly additional discrete CME component (i.e., the second one). The third row focuses on velocity distribution within the QS region. Figure~\ref{fig:disk_v} yields robust results for acquiring CME spectra across a vast velocity range.

\begin{figure}[ht!]
\plotone{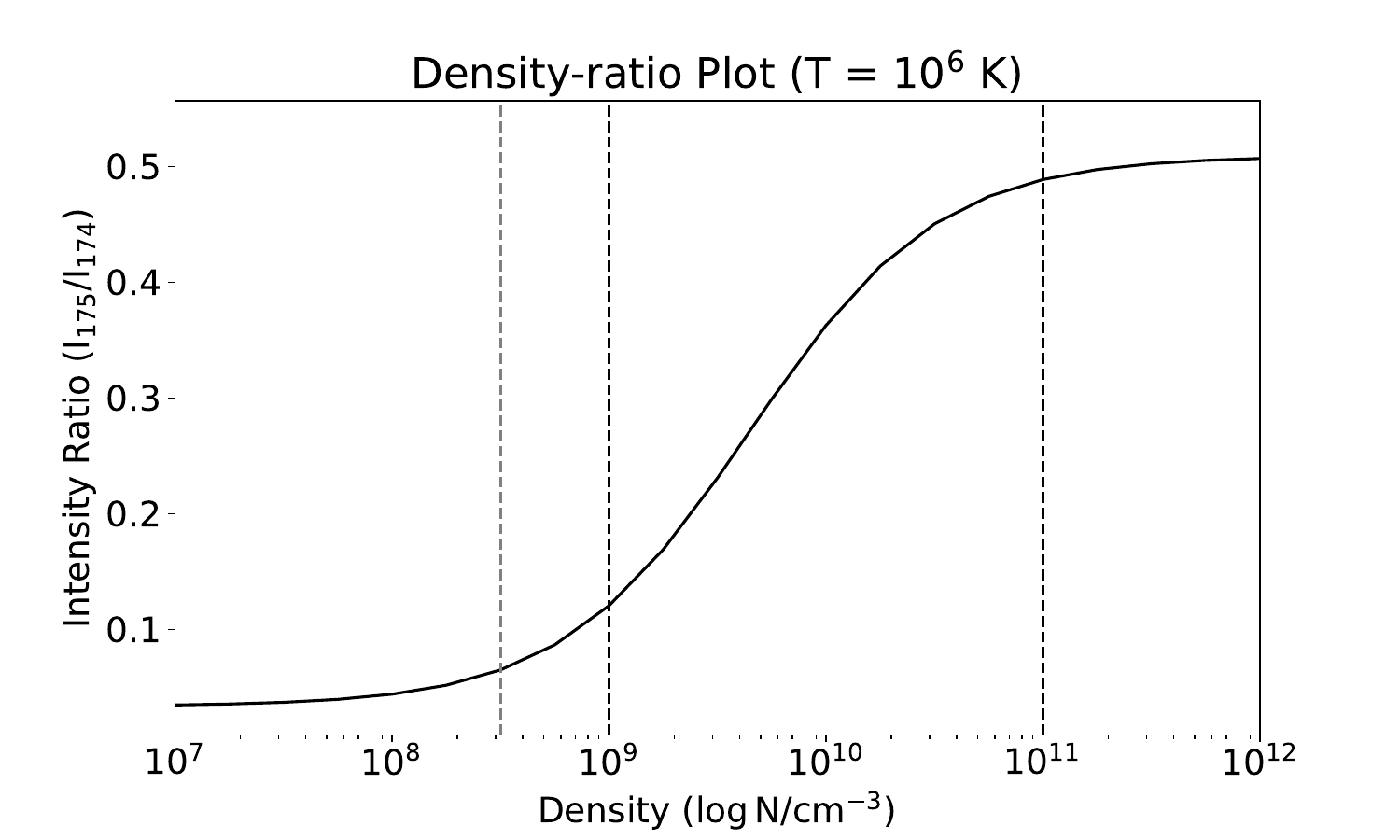}
\caption{Intensity ratio of the Fe~{\sc{x}} 175.26 $\rm \AA$ and Fe~{\sc{x}} 174.53 $\rm \AA$ lines as a function of electron density, calculated at a temperature of $\rm 10^6\,K$. The two black vertical lines represent the density-sensitive range ($9<\log\,N/\text{cm}^{-3}<11$), as shown in CHIANTI database and \citet{DelZanna2018}. The grey vertical line corresponds to a density of $\log\,N/\text{cm}^{-3} = 8.5$, which serves as a density threshold applied in Figure~\ref{fig:disk_i_n}. This threshold is selected because the curve becomes flat around this point, leading to large uncertainties in diagnostics measurements for densities below this value.}
\label{fig:density_ratio}
\end{figure}

\begin{figure}[ht!]
\plotone{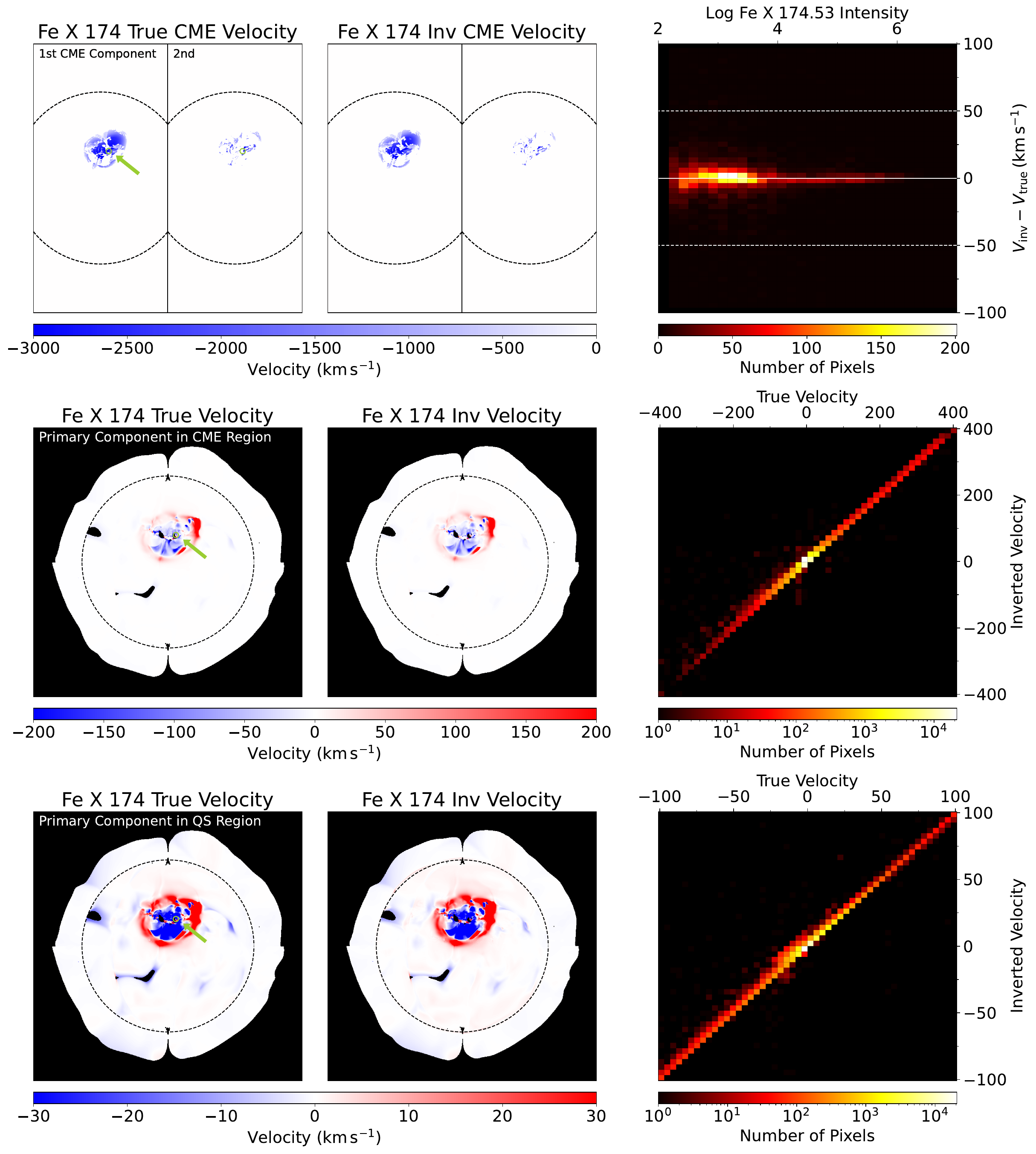}
\caption{Comparison between the ground truth (left column) and inverted results (middle column) for the Doppler shift of Fe~{\sc{x}} 174.53 $\rm \AA$ by JPDFs (right column). The green circle and arrow mark the corresponding on-disk pixel in Figure~\ref{fig:eg_sp}. The first and second CME components (top row) are shown in left and right subpanel (marked by ``1st CME Component'' and ``2nd''), respectively. Primary components with higher-velocity (lower-velocity) colorbar in middle row (bottom row) show the distribution of relatively low velocity from -400 to 400 $\rm km\,s^{-1}$ (from -100 to 100 $\rm km\,s^{-1}$) in CME region. {The second row shows a belt-like structure in the CME region with lower redshifts. We consider these as primary components because the profile can be fitted using a double Gaussian that has some overlap between the primary (rest) component and the redshift component. The third row shows the Doppler velocity distribution within the QS region, where slight redshifts are observed in certain region. These velocity ranges are presented in the JPDFs in the second and third row, demostrating strong agreement between the ground truth and inverted results.}  The JPDF in the first row (white line shows $\rm \pm\,50\,km\,s^{-1}$ uncertainties calculated from resolving power of $\sim$2000) shows the difference between the ground truth and the inversion results as a function of logarithmic Fe~{\sc{x}} 174.53 $\rm \AA$ intensity with threshold of $\rm S/N = 10$  and $I_{\text{CME}} = I_{\text{max}}^{l}/20$.  }
\label{fig:disk_v}
\end{figure}

\begin{figure}[ht!]
\plotone{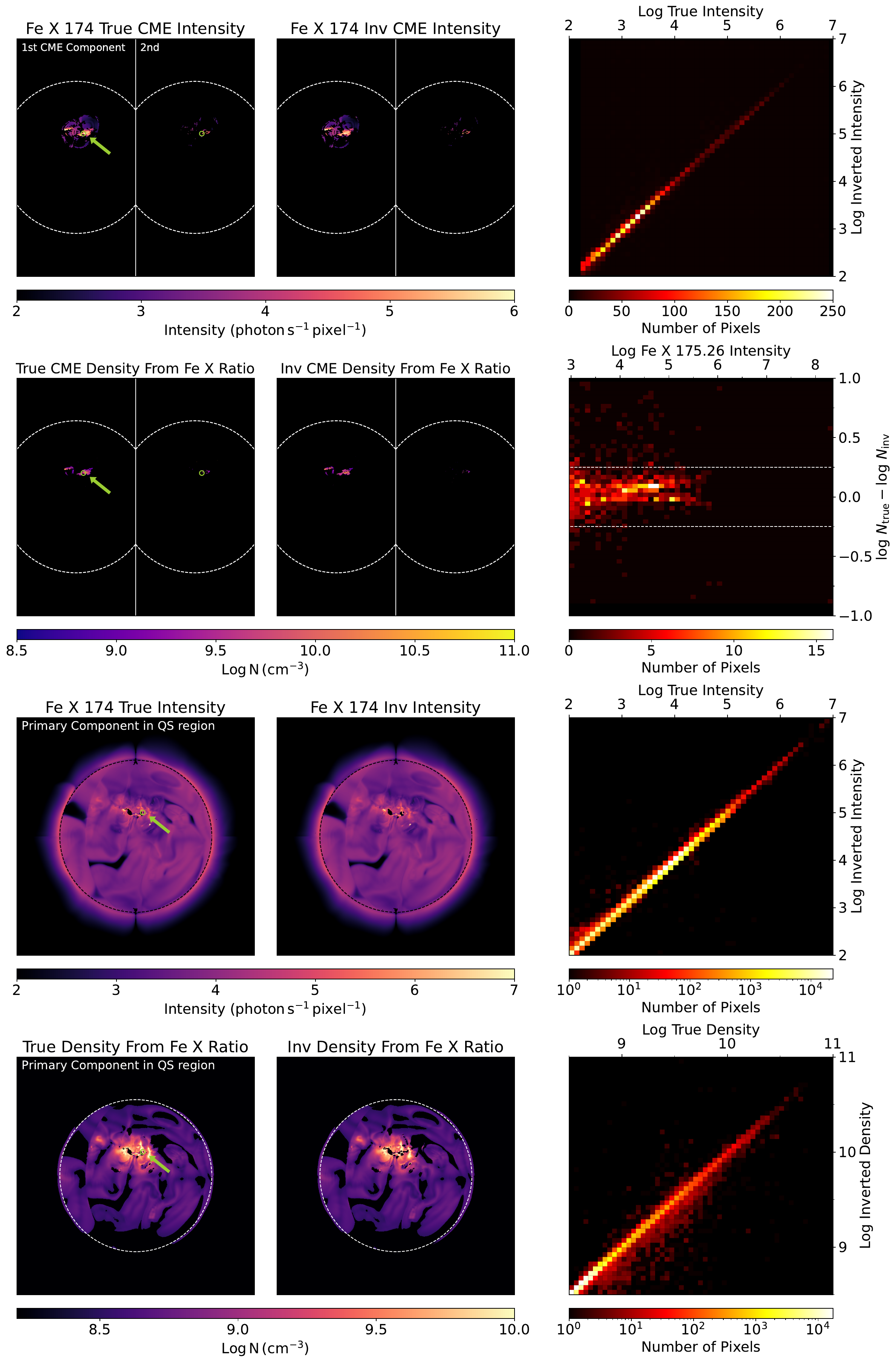}
\caption{Similar to Figure~\ref{fig:disk_v} but the JPDF in the second row is with intensity threshold of $\rm S/N = 30$ of Fe~{\sc{x}} 175.26 $\rm \AA$ intensity and density threshold of $\log{N}/\text{cm}^{-3}= 8.5$.  The white dashed  line in the second row shows 25\% uncertainties.}
\label{fig:disk_i_n}
\end{figure}

The density diagnostics for an on-disk CME can be derived from a density-sensitive line pair---Fe~{\sc{x}} 174.53/175.26 $\rm \AA$. Figure~\ref{fig:disk_i_n} shows the intensity maps of Fe~{\sc{x}} 174.53 $\rm \AA$ and the density maps for both the CME components and the primary component using the same threshold of Figure~\ref{fig:disk_v}. The top-right region of the first CME components in the intensities of Fe~{\sc{x}} 174.53 $\rm \AA$ (first row) exhibits relatively low intensities with high velocities (see Figure~\ref{fig:disk_v} for the maps with large velocities). Conversely, certain pixels in the region below shows a substantial intensity for both the first and the second CME components, which possess spectra similar with those of Figure~\ref{fig:eg_sp}. In pixels with strong intensities, the density for the second CME components can be accurately determined, e.g., the consistent bright patterns shown in the first row and the second row. However, in other regions of the density maps for the CME components, the {number of pixels with detectable densities} is significantly lower than the corresponding pixels in the intensity maps, attributed to the moderate performance of the intensities of Fe~{\sc{x}} 175.26 $\rm \AA$. {This moderate performance is attributed to the relatively low intensity of  Fe~{\sc{x}} 175.26 $\rm \AA$ compared to other primary lines (as shown in Figure~\ref{fig:eg_sp}). This situation is further exacerbated for the fast CME components of this spectral line with high blueshifts. These components are often blended with other spectral lines, particularly the primary components of  Fe~{\sc{x}} 174.53 $\rm \AA$.} To optimize this, we applied an intensity threshold of $\rm S/N=30$ for Fe~{\sc{x}} 175.26 $\rm \AA$ intensity. A density threshold of  $\log{N}/\text{cm}^{-3}= 8.5$ is also utilized, as the sensitivity range of this line pair is $\log{N}/\text{cm}^{-3}= 9\,-\,11$, showing considerable uncertainties in density measurements below $\log{N}/\text{cm}^{-3}= 9$, particularly for densities lower than $\log{N}/\text{cm}^{-3}= 8.5$, {where the density-intensity ratio curves become nearly flat (as shown in Figure~\ref{fig:density_ratio}).} Significant intensity decrease (lower than the given $\rm S/N$) within the CME region is distinctly observed in the intensity maps (third row) combined with the velocity maps for the first CME components in Figure~\ref{fig:disk_v}, indicating a significant intensity decrease (maybe dimming) at the center of CME region but with high blueshift. In summary, we successfully obtained the density maps for the CME components for strong-signal regions within an uncertainty of $25\%$, shown in the second row of Figure~\ref{fig:disk_i_n}. The density map for the primary component (fourth row) can be accurately derived, but this is limited to the on-disk regions due to the low intensity of Fe~{\sc{x}} 175.26 $\rm \AA$ intensity, density sensitivity of the line pair, and various velocity components in the off-limb region.

\subsection{Off-limb CME Scenario}\label{sec:offlimb}
\begin{figure}[ht!]
\plotone{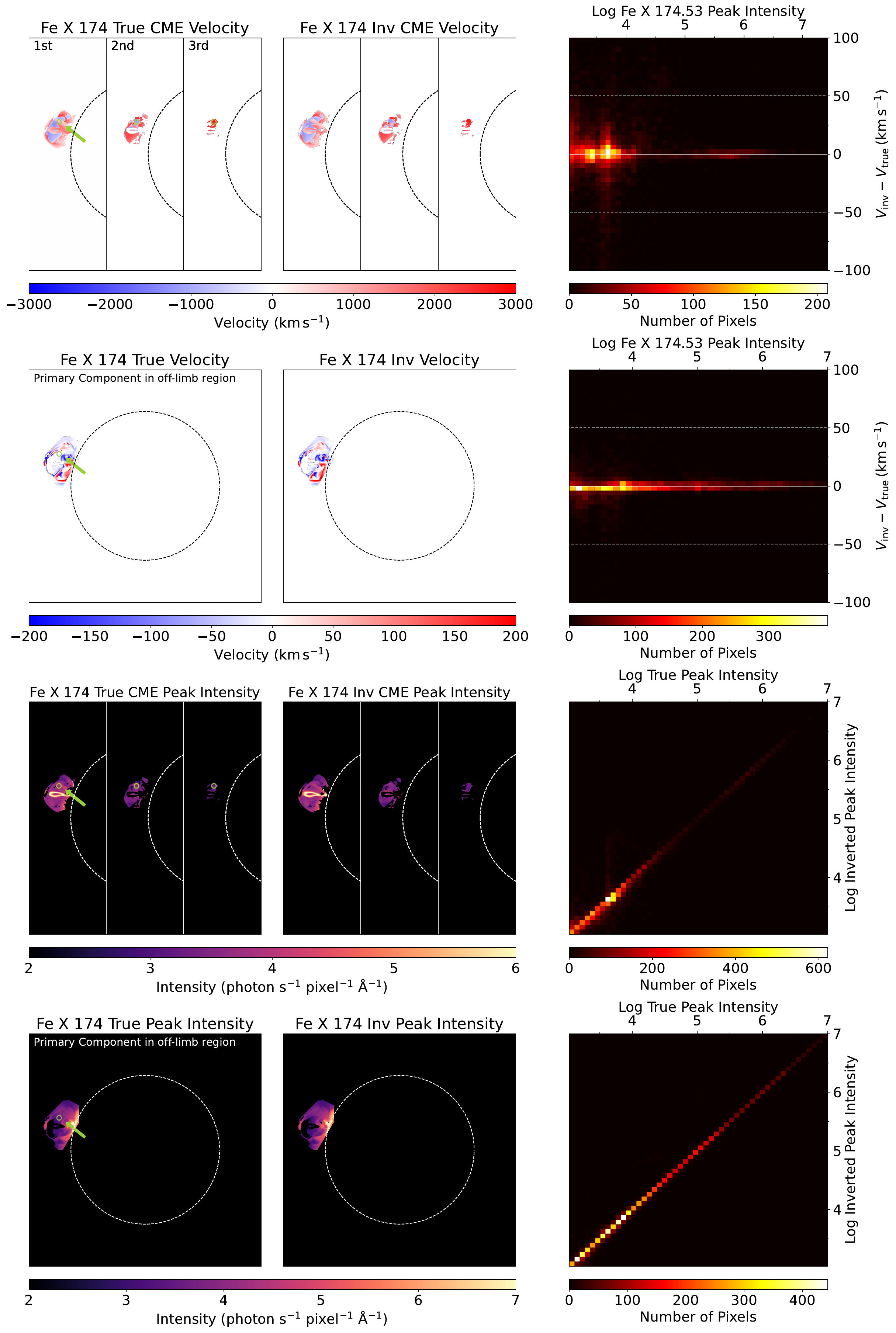}
\caption{Similar to Figure~\ref{fig:disk_i_n} but for the off-limb Doppler shift (1st row) and peak intensity (3rd row) of the first, second and third CME components, and for the off-limb Doppler shift (2nd row) and intensity (4th row) for primary components. The green circle and arrow mark the corresponding off-limb pixel in Figure~\ref{fig:eg_sp}. JPDFs in the first and the second row are similar to Figure~\ref{fig:disk_v} but with a peak intensity threshold of $\rm S/N\sim32.9$ (i.e., $\log{3.03}$ for logarithmic photon counts)}
\label{fig:limb_v_i}
\end{figure}

An off-limb CME, obtained by varying viewing angle of the same frame, can be also detected using the multi-slit EUV spectroscopy. However, density and line width diagnostics were not considered while we used peak intensity instead. This {inability} is attributed to the presence of various velocity components in the off-limb scenario, complicating {the inversion process and} the line width diagnostics. The peak intensity $I_{\text{p}}$ is calculated by the total line intensity $I = I_{\text{p}}\sqrt{\pi}\Delta\lambda$, where $\Delta\lambda$ is the line width ($\rm 1/e$ width). {It it important to note that this formula is only for calculating the S/N threshold in off-limb scenarios, where the total intensity corresponds to the S/N threshold of 10 and the line width is an average value that includes instrumental broadening, thermal broadening with coronal temperature, and non-thermal broadening of $\sim$$\rm 15\,km\,s^{-1}$. In contrast, the peak intensity maps are derived directly from the y axis values of its spectrum. }The numbers of the CME components in the off-limb scenario (three discrete components shown in Figure~\ref{fig:limb_v_i}) slightly differs from those in the on-disk scenario (two discrete components) for this case. Similar to {last (second) CME components in }the on-disk scenario, the occurrence of the last (third) components {in the off-limb scenario} is relatively infrequent (shown in the subpanel in the first row in Figure~\ref{fig:limb_v_i}). Considering that CMEs have a bubble-like structure that expands outward, it is reasonable to observe corresponding patterns, e.g., the upside-down triangle pattern shown in the first row in the Figure~\ref{fig:limb_v_i}. For this pattern, the first (blueshift) and the second (redshift) CME components perform similarly with discrete redshift and blueshift CME components shown in the bottom panel in Figure~\ref{fig:eg_sp}. The flux rope is also clearly visible in the velocity maps (first row) and the intensity maps (third row), exhibiting an overall redshift with weak primary components and the absence of the second and the third CME components. This is because the propagation direction slightly orients away from the observers (arbitrarily selected viewing angle for the off-limb scenario in the model). The maps for primary components (second row and fourth row) show missing pixels of the flux rope for the same reason. The CME components are sorted based on intensity, indicating that the velocity maps of  Fe~{\sc{x}} 174.53 $\rm \AA$ for the first CME components show the velocity distribution with the strongest intensity, rather than the strongest Doppler shift. For example, in the triangle pattern with blueshift (redshift) for the first (second) CME components, the redshift pattern has a larger offset velocity than the blueshift.

\begin{figure}[ht!]
\plotone{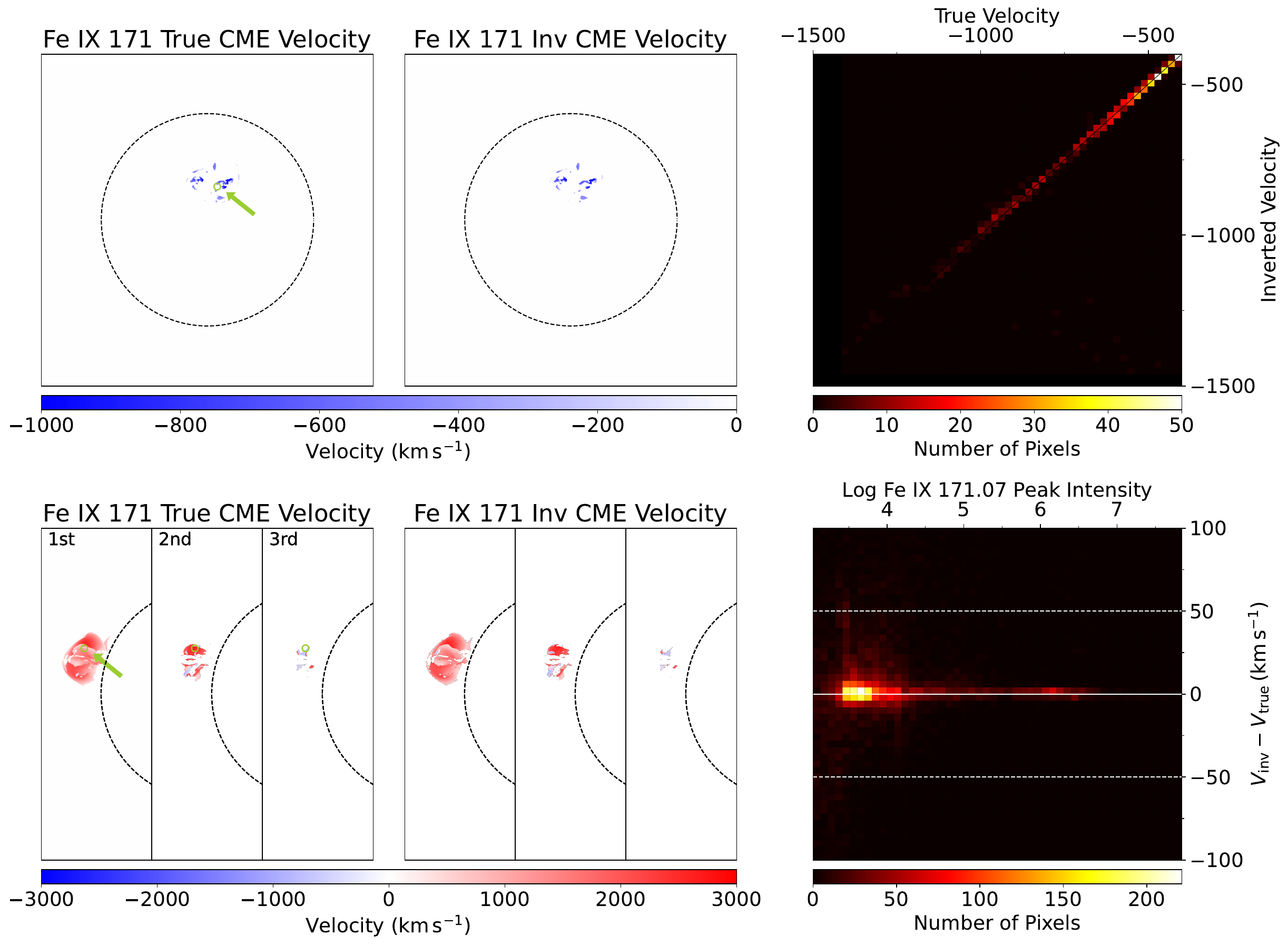}
\caption{Similar to Figure~\ref{fig:disk_v} and \ref{fig:limb_v_i} but for the Doppler shift for Fe~{\sc{ix}} 171.07 $\rm \AA$ in the on-disk (top row) and the off-limb scenarios (bottom row), respectively.}
\label{fig:171v}
\end{figure}

The detection of Fe~{\sc{ix}} 171.07 $\rm \AA$ encounters challenges in both the on-disk and the off-limb scenarios because of the aforementioned sharp profile of the effective area around 171 Å. In the on-disk scenario, the number of detectable pixels is significantly lower compared to Fe~{\sc{x}} 174.53 $\rm \AA$ (with blueshifts only below about $\rm-1500\,km\,s^{-1}$ shown in the first row in Figure~\ref{fig:171v}). Most detectable velocity pixels are below a blueshift of about $\rm-1000\,km\,s^{-1}$ while only a small fraction exceeding this velocity. The situation would be worse in practice because of the instrumental uncertainties at the edge of the effective area. In principle, the velocity patterns for Fe~{\sc{ix}} 171.07 $\rm \AA$ and Fe~{\sc{x}} 174.53 $\rm \AA$ should be similar. However, the velocity patterns for Fe~{\sc{ix}} 171.07 $\rm \AA$ (shown in the second row in Figure~\ref{fig:171v}) predominantly exhibit redshifts, whereas those for Fe~{\sc{x}} 174.53 $\rm \AA$ (shown in Figure~\ref{fig:limb_v_i}) show a more balanced distribution of redshifts and blueshifts. This discrenpancy results from the combined effects of the effective area profile and our sorting criterion for the CME components by intensity. Some patterns with strong blueshifts and significant decreases in intensity may be classified as redshift patterns or may even disappear due to their low intensities (lower than aforementioned intensity threshold). The different patterns for these two spectral lines can be interpreted as a result of a ``replacement''. For example, initially, they exhibit similar patterns when the effective area and sorting criteria are not taken into account. However, once these factors are considered, the upside-down triangle blueshift patterns for the first CME components (shown in the first row in Figure~\ref{fig:limb_v_i}) would be classified as the redshift patterns (shown in the second row in Figure~\ref{fig:171v}). Extending this analysis to other pixels would result in a majority of redshifts in the velocity maps for Fe~{\sc{ix}} 171.07 $\rm \AA$. Although Fe~{\sc{ix}} 171.07 $\rm \AA$ is less suited for observing fast CME spectra, it provides strong signals for CH observations (see Table~\ref{tab:primary lines}) compared with other primary lines, and offers different temperature information ($\log{T}/\text{K}= 5.9$), enriching the plasma diagnostics and broadening observational targets.

\section{Discussion and Conclusions} \label{sec:conclusion}
In this paper, we have introduced a scheme for global observations of the solar corona, using a multi-slit EUV spectroscopy combined with a new application of {our} decomposition technique. We focus on the detectability of CME spectra for both the on-disk and the off-limb scenarios, especially addressing an ambiguous parameter---velocity. We have incorporated four primary lines (Fe~{\sc{x}} 174.53 $\rm \AA$, Fe~{\sc{x}} 175.26 $\rm \AA$, Fe~{\sc{x}} 177.24 $\rm \AA$, and Fe~{\sc{ix}} 171.07 $\rm \AA$) to facilitate global plasma diagnostics, including Doppler shift and density diagnostics. We focused on Doppler shift diagnostic, primarily using Fe~{\sc{x}} 174.53 $\rm \AA$, for a fast CME (reaching up to a velocity of around $\rm-3000\,km\,s^{-1}$). In the on-disk scenario, we have successfully identified discrete and unseparated CME components, and obtained their partial density maps of a CME. We found significantly fewer detectable pixels for Fe~{\sc{ix}} 171.07 $\rm \AA$ compared to those for Fe~{\sc{x}} 174.53 $\rm \AA$ due to the sharp decrease of the effective area, limiting detectable velocities to below about $\rm-1500\,km\,s^{-1}$. In the off-limb scenario, the complicated velocity components hinder the density diagnostic because of the poor performance of Fe~{\sc{x}} 175.26 $\rm \AA$, which has intrinsically low intensity, and also hamper line width inversion for Fe~{\sc{x}} 174.53 $\rm \AA$. Unusual redshift-dominant patterns were found in the velocity maps for Fe~{\sc{ix}} 171.07 $\rm \AA$ as a result of the sharp profile at the edge of the effective area combined with our sorting criteria for multiple discrete CME components. The comparison between the ground truth and the inverted results represents a good agreement for plasma parameters for both CME and primary components in both scenarios. Note that Fe~{\sc{ix}} 171.07 $\rm \AA$ is expected perform well in scenarios with plasma moving at relatively low velocities, such as slow CMEs and CHs.

Our proposed scheme with multi-slit design and strategies for identifying CME components balances detector size and wavelength range. A clean wavelength range with at least two strong spectral lines from the same ion is essential. The wavelength difference of a density-sensitive line pair should be carefully chosen because a close line pair (e.g., Fe~{\sc{x}} 174.53/175.26 $\rm \AA$) could affect each other during CMEs, particularly affecting the weaker one, whereas a distant one (e.g., Fe~{\sc{xii}} 195.12/186.89 $\rm \AA$), would require a larger detector. A five-slit design represents the minimum necessary to meet scientific goals. Although additional slits could be considered, it would introduce more potential overlap and increase calculation resources of inversion. The ambiguity of velocity components during CMEs, especially in off-limb scenarios, complicates spectral line analysis. Therefore, we adopted a new application of {our} decomposition technique, decomposing velocity ambiguity instead of slit number (see \citet{Cheung2019,DePontieu2020,Chan2024} as examples for the decomposition of slit number). While a third (fourth) CME component in the on-disk (off-limb) scenario might exist, they are likely to be few in number or too weak in intensity. Note that the multiple CME velocity components indicate the presence of several distinct plasma with different velocities along the LOS column. However, a detailed discussion of their physical implications is beyond the scope of this paper. Additionally, future efforts of {our} decomposition process should include the line width. {In this proposed instrument, the primary focus is on the Doppler shift of a CME, while the line width is of relatively minor importance. Furthermore, in our decomposition technique (likely in others as well), the diagnostics for line width are the least accurate, with accuracy further deteriorating during a CME event.}

{We utilized a snapshot from a CME onset MHD simulation, although the nature of this event is not the primary focus of our current investigation.  To clarify,  CME onsets are complicated and diverse phenomena that can involve multiple velocity components, encompassing not only the mass ejected during the eruption but also large-scale waves and shocks. Notably, off-limb velocity maps indicating flows of a few thousand km/s may be indicative of large-scale waves or shocks. The proposed instrument concept has the potential to not only detect the nature mentioned in this work but also to identify large-scale waves and shocks, thereby broadening its capabilities.}

There is an inherent trade-off between temperature coverage and the decomposition process, e.g., a wavelength range containing ions with a broad range of formation temperatures, could lead to spectral crowding and overlap, particularly during CMEs. Flare lines were not included in this study or in \citet{Chan2024}, future proposed scheme could consider incorporating them. Also, magnetic field strength, which is a key factor in solar activity and affects the line intensities, was not addressed in this paper. The phenomenon of magnetic-ﬁeld-induced transition (MIT) has been shown promising for measuring coronal magnetic ﬁeld strength \citep[][]{Li2015,Li2016,Chen2021,Martinez2022,Chen2023}. In the future, a scheme containing MIT lines with multi-slit EUV spectroscopy could be carefully investigated. Other advanced machine learning techniques could also enhance the efficiency of identification of CME components, {e.g., the deep neural network}. Using existing or adopting a context imager in 174 Å channel in practical observations could support {our} decomposition process as a guide for the spectrograph \citep[e.g.,][]{Winebarger2019,DePontieu2020} because they could offer additional constraints for {our} decomposition process(i.e., the integrated spectral intensity). A context imager can provide POS velocities combined with LOS velocity (obtained by EUV spectroscopy) for generating velocity vectors of CMEs in the low corona, improving the accuracy of space weather forecasting.

\begin{acknowledgments}
We thank the reviewer for their constructive suggestions.
We also thank Lucie Green and Zihao Yang for insightful discussions of the work.
This research work is supported by the National Key R\&D 78
Program of China (2021YFA1600500, 2022YFF0503800, 2021YFA0718600), the Youth Innovation Promotion Association CAS (2023061) and the National Natural Science Foundation of China (NSFC, Grant Nos. 12103066, 12073004).
T.T. acknowledges support from NSF award ICER-1854790 and from NASA awards 80NSSC20K1274 and 80NSSC24K1108.
CHIANTI is a collaborative project involving George Mason University, the University of Michigan (USA), University of Cambridge (UK), and NASA Goddard Space Flight Center (USA).
\end{acknowledgments}

\vspace{5mm}
\bibliography{bibfile}{}
\bibliographystyle{aasjournal}

\begin{table}[h!]
  \begin{center}
    \caption{Primary lines used for global plasma diagnostics.}
    \begin{tabular}{lcccccccccc}
    \hline\hline
    \multirow{1}{*}{Ion and} & \multirow{1}{*}{Formation} & \multicolumn{7}{c}{Expected Signal ($\rm ph\, s^{-1}\, pixel^{-1}$)}   \\
    \cline{3-9}
    \multirow{1}{*}{Wavelength (\AA)} & \multirow{1}{*}{Temperature ($\log\,T/\text{K}$)} && CH && QS && AR \\
    \hline
    Fe~{\sc{ix}} 171.07               & 5.9 && 582.3  && 2310.5  && 21414.7   \\
    Fe~{\sc{x}} 174.53                & 6.0 && 151.5  && 1770.0  && 22725.9   \\
    Fe~{\sc{x}} 175.26                & 6.0 && 18.1  && 196.2  && 2473.5 \\
    Fe~{\sc{x}} 177.24                & 6.0 && 46.7   && 544.6  && 6990.6   \\
    \hline
    \end{tabular}
    \tablecomments{Four primary lines and their corresponding formation temperature are shown in the first and second columns, respectively. A density-sensitive line pair is included (Fe~{\sc{x}} 174.53 $\rm \AA$ and Fe~{\sc{x}} 175.26 $\rm \AA$). The expected signals (third column) for four primary lines are calculated by three standard CHIANTI DEMs (CHs, QS, and ARs), assuming a density of $10^9\,\text{cm}^{-3}$. {The displayed signals result from the convolution of the effective area (shown in Figure~\ref{fig:effective area}), and are given in units of $\rm ph\,s^{-1}\,pixel^{-1}$}.}
    \label{tab:primary lines}
    \end{center}
\end{table}

\end{CJK*}
\end{document}